\documentclass[conference]{IEEEtran}

\ifCLASSINFOpdf
  \usepackage[pdftex]{graphicx}
 
\else
  
\fi

\usepackage{color,soul}

\usepackage{multirow}
\usepackage{color, colortbl}
\definecolor{Gray}{gray}{0.9}

\usepackage{url}

\begin{document}

\title{Are Smart Home Devices Abandoning IPV Victims?}

\author{\IEEEauthorblockN{Ahmed Alshehri}
\IEEEauthorblockA{Accenture Labs\\ Arlington, VA\\}
alshehri@mines.edu
\and
\IEEEauthorblockN{Malek Ben Salem}
\IEEEauthorblockA{Accenture Labs\\ Arlington, VA\\}
malek.ben.salem@accenture.com
\and
\IEEEauthorblockN{Lei Ding}
\IEEEauthorblockA{American University\\ Washington, D.C.\\}
ding@american.edu
}

\maketitle
\thispagestyle{plain}
\pagestyle{plain}
\begin{abstract}
\label{abstract}
Smart home devices have brought us many benefits such as advanced security, convenience, and entertainment. However, these devices also have made unintended consequences like giving ultimate power for devices' owners over their intimate partners in the same household which might lead to tech-facilitated domestic abuse (tech-abuse) as recent research has shown. 
In this paper, we systematize findings on tech-abuse in smart homes.
We show that domestic abuse and Intimate Partner Violence (IPV) in smart homes is more effective and less risky for abusers. Victims find it more harmful and more challenging to protect themselves from.
We articulate a comprehensive analysis of all the phases of abuse in smart homes and categorize risks and needs in each phase. 
Technical analysis of current smart home technologies is conducted to shed light upon their limitations.
We also summarize recent recommendations to combat tech-abuse in smart homes and focus on their  potentials and shortcomings.
Unsurprisingly, we find that many recommendations conflict with each other due to a lack of understanding of phases of abuse in smart homes.
Desirable properties to design abuse-resistant smart home devices are proposed for all the phases of abuse.
The research community benefits from our analysis and recommendations to move forward with a focus on filling the blind spots of existing smart home devices' safety measures and building appropriate safety measures that consider tech-abuse threats in smart homes.

\end{abstract}


%
\IEEEpeerreviewmaketitle

\section{Introduction}
\label{introduction}

Smart home devices have been adopted at a high rate in recent years. 
There is at least one smart home device in 69\% of the houses in the U.S.~\cite{SmartHome_Stat_majority_69}. 
Three in five Americans own smart home devices for security purposes in their houses~\cite{SmartHome_Security_stats}.
Many smart home devices can be programmed to do a variety of tasks. 
For example, Amazon \textit{Alexa} allows users to create their own skills such as ``start relaxing music when I get home''~\cite{Alexa_Skills}. 
This sounds entertaining for most people. 
Nevertheless, this same feature can be turned to be a great tool for eavesdropping~\cite{Alexa_Skills_Survillance}.
Recent articles in the news have reported cases of domestic abuse using smart home devices~\cite{Wired-smartHomes_abuse, News_SmartHome_Abuse}.
Smart home devices are normally designed for homes where trust is established without any dispute.
This trust assumption is not realistic as IPV cases are common globally.
According to the United Nations' 2019 global study on homicide, one in three women has been killed by their partners~\cite{UN_Report_homicide}.
The same report shows that two in three women are more likely to be killed at their home by a relative.

Technology-facilitated domestic abuse (tech-abuse for short) is not new.
Abusers have been using technologies like smartphones, personal computers, and GPS trackers to improve their methods of abuse. The emergence of smart home devices has just further advanced the methods that abusers would use~\cite{SmartHome_Abuse_Category_1_paper1}. 
Smart home technologies give abusers a great tool to carry out destructive activities with less effort and lower risks compared to traditional methods of abuse~\cite{SmartHome_Abuse_Category_1_paper2}. 
Still, a thorough analysis of tech-abuse in smart homes has not been conducted.

Consequently, many research projects have reacted to tech-abuse in smart homes with ways to protect the victims of Intimate Partner Violence (IPV) such as the work in~\cite{SmartHome_Abuse_Category_2_paper1, SmartHome_Abuse_Category_2_paper2, SmartHome_Abuse_Category_2_paper3}. 
Some of these projects have analyzed victims' perspectives in IPV cases through user studies and workshops. The proposed solutions to help the victims are either generic like advising IPV support services to involve cybersecurity experts for more appropriate planning or specific to one aspect of the system such as the focus of~\cite{SmartHome_Abuse_Category_2_paper3} on the usability of smart home devices. Both approaches are good to a certain extent.
Still, comprehensive analysis and solutions are missing since the issue of tech-abuse in smart homes is still emerging. 

In this paper, we develop a unified analytical framework based on characteristics of smart homes, abusers' common attributes, and possible capabilities particular to smart homes built with emerging technologies.
Our framework consists of four phases of abuse from the victim's perspective in IPV.
For each phase, we explore the goals for each stakeholder including abusers, victims, device vendors, and support services. 
Then, we review the literature with some key insights and break down security and safety challenges and needs for each phase.
The research on ensuring human safety in smart homes is somewhat less well developed.
\begin{table*}[!t]
	\centering
	\caption{Scenario-based comparison of abuse methods. We take \textit{stalking} as an abusive action.} 
	\vspace{-8pt}
	\label{tab:techs}
	\scalebox{0.9}{
	\begin{tabular}{|l|l|}
	    \hline
	    \rowcolor{Gray}
		\textbf{Abuse type}    &  \textbf{What abusers can do}        \\ \hline
		
	   \multirow{2}{*}{Traditional abuse}&High effort:\textit{ Abusers need to spend time following the victim, be physically close, and abusive action is not immediate.} \\
         &High risk: \textit{Abusers might face resistance by victims to fight back physically, and other people might see or hear the abusive action.}\\&High benefit: \textit{Abusers can assert their power over victims.}\\ \hline
          \multirow{2}{*}{Tech-abuse}&Mid effort: \textit{Abusers  need to either be physically close to victims' devices or know the credentials, and actions are not immediate.}  \\
         &High risk: \textit{The same risk in traditional abuse applies here, and some recent defenses for victims can hold abusers accountable~\cite{Tech-abuse-smartphones-computers}.} \\&Low benefit: \textit{Restricted power for abusers as victims need to initiate activities on their phones or PCs first.} \\ \hline
          \multirow{2}{*}{SHOT}&Low effort: \textit{Abusers own the devices of the smart homes which give them full access remotely and immediately.}  \\
         &Low risk: \textit{Abusers can control their smart homes remotely and are able to delete usage history.} \\&High benefit: \textit{Abusers can control ambient surroundings of victims, and are able to discretely monitor the house.} \\ \hline
	\end{tabular}
	}
	\label{abuse_comparison}
	\vspace{-15pt}
\end{table*}

Then, we analyze the technical causes of why smart home devices empower abusers in IPV, limit victims' abilities to protect themselves, and challenge support services to provide help.
Lack of  diversity  in the designing teams, poor usability of the devices, inadequate modeling of the IPV threat in smart homes, and unreliable system functions are the main  reasons.
We then review proposed solutions by recent work to show their strengths and weaknesses as well as future directions along with desirable properties for abuse-resistant smart home devices.
To the best of our knowledge, we are the first to design a specific framework for phases of abuse in smart homes, and break down security challenges to each phase so that the research community can target the limitations of current research, and focus on a comprehensive solution that works for all the phases of abuse in smart homes.

\textbf{Our contributions:}
\begin{itemize}

    \item We introduce a unified analytical framework to allow structured analysis of tech-abuse in smart homes. This framework is developed from previous efforts by considering the entire lifecycle of tech-abuse, instead of smart home technologies in isolation. (Section \ref{phases})
    \item We break down the tech-abuse problems into smaller privacy and security (safety) problems in each phase. This simplifies the problem and aids in addressing it at each phase. (Section \ref{phases}) 
    \item We observe that smart homes empower abusers and limit IPV victims' ability to protect themselves. (Section \ref{technical_causes})
    \item We summarize solutions to combat tech-abuse in smart homes to show their strengths and weaknesses in order to enhance existing ideas and guide the research community to cover the blind spots of current research. (Section \ref{summary_solutions})
    \item We systematize desirable properties to make smart home devices abuse-resistant. (Section \ref{ourProposals})
\end{itemize}


\section{Background}
\label{background}
\subsection{Background}

According to the United States Department of Justice Office on Violence Against Women, domestic violence is a pattern of abusive behavior in any relationship that is used by one partner to gain or maintain control over another intimate partner~\cite{DoJ_OVW_Domestic_Definition}. 
There are three types of abuse: 
\begin{itemize}
	\item \textbf{Traditional abuse}: Any abusive actions that do not use technology, 
	\item \textbf{Tech-abuse}: Any abusive actions using technologies such as smartphones, personal computers, or social media websites, and 
	\item \textbf{Tech-abuse in smart homes}: Any abusive actions that use smart home devices. 
\end{itemize}

To prevent ambiguity, we refer to tech-abuse in smart homes as Smart HOme facilitated Tech-abuse (\textit{\textbf{SHOT}}).

Many recent research projects on domestic abuse in smart homes were user studies. The studies qualitatively analyze victim's needs in smart homes or front-liners' needs of support services to combat abusers in IPV~\cite{SmartHome_Abuse_Category_1_paper1,SmartHome_Abuse_Category_2_paper1, SmartHome_Abuse_Category_2_paper2, SmartHome_Abuse_Category_1_paper2}. 
Other researchers partially considered SHOT without analyzing its corresponding specific abuse phases.
Havron et al. propose a solution similar to a medical clinic, where an IPV victim can reach out to support services and an expert can provide personalized help~\cite{tech-abuse-helpVictims_2}.
Kotz et al. discuss unique challenges in ensuring human safety in smart environments and encourage designers to consider these challenges while developing products~\cite{challenges_safety_smart_Environments}.
Nuttall et al. propose five design principles that vendors should consider when developing new technologies to prevent coercive control for abusers~\cite{IBM_tech_abuse_design}. 
These research efforts focus on tech-abuse generally and we argue that tech-abuse in smart homes is unique and calls for specific considerations.
In short, there has not been a thorough analysis of specific phases of abuse in smart homes, technical reasons for empowering abusers, and concrete breakdowns of security and safety issues of SHOT.
Safety in this paper means that smart home security and privacy measures fail and personal safety of users becomes at risk.

\subsection{Differences between Tech-abuse and SHOT}
Tech-based abuse is when abusers use technology such as smartphones or computers to carry out harmful actions to their partners. 
There are similar characteristics of this kind of abuse. 
Devices are owned by the victim but might be accessed by the abuser.
Threats lie around what the victim might do in their gadgets such as calls made, posts on social media, or usage of some apps. 
The research community has reasonably offered Privacy Enhancing Technologies (PET) to victims to protect themselves~\cite{tech-abuse-helpVictims_1,tech-abuse-helpVictims_2}. 
In short, smartphones and personal computers have been the subject of research in tech-abuse. This has led to many defense mechanisms. 
\begin{figure*}[t]
\centerline{
\includegraphics[scale=.5]{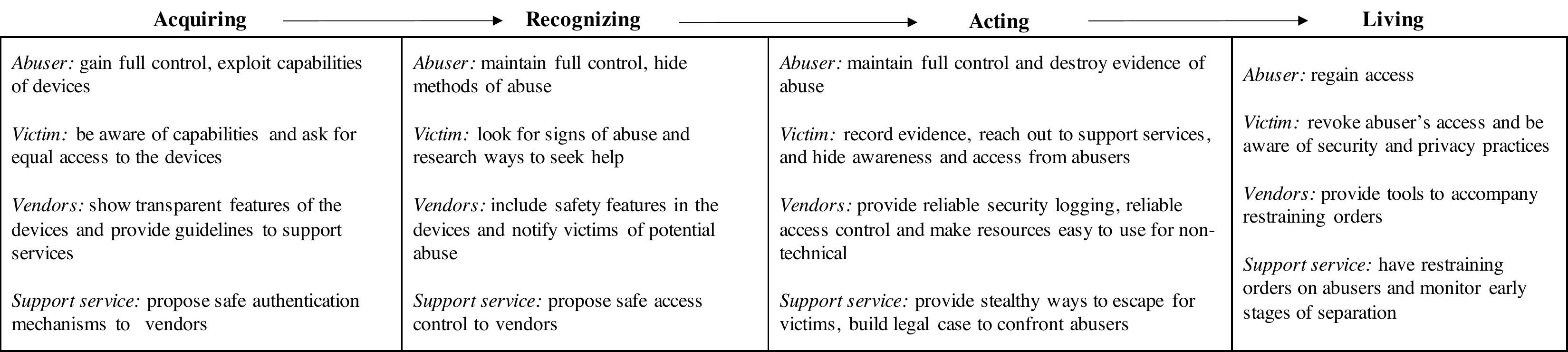}}
\vspace{-0.4cm}
\caption{Our framework of phases of SHOT with the corresponding goals for each entity involved.}
\label{phases_figure}
\vspace{-17pt}
\end{figure*}

Nevertheless, tech-abuse in smart homes (SHOT) is relatively new. 
This calls for more research to explore the technical and social complexities as well as the design issues. 
Abusers' abilities to harm their intimate partners have advanced greatly with smart home devices. Abusers can now use effective, efficient, and untraceable methods to carry out stalking, gaslighting, and other abusive actions~\cite{SmartHome_Abuse_Category_1_paper1}. 
Also, devices in smart homes are not necessarily owned by victims. Rather, they are largely owned and controlled by abusers~\cite{SmartHome_Abuse_Category_1_paper2}. 
These devices can modify the ambient environment and unlock the door of the house. 
Simply, they can impact the physical surroundings which is more damaging than just dealing with digital data. 

Table \ref{abuse_comparison} demonstrates the differences between the three types of abuse. We consider three metrics to evaluate the abuse methods: the required effort to carry out the abuse, the risk taken by abusers, and the benefit to the abusers.
Estimating effort takes into account the time spent carrying out the abuse, the proximity of the abuser to the victim, and the speed of the action (whether immediate or not).
Risk is approximately measured by how victims can defend themselves which will impact abusers and hold them accountable.
In the formal definition of domestic abuse, abusers seek to control their intimate partners for a variety of reasons.
In Table  \ref{abuse_comparison}, we estimate the benefits to abusers by looking at how much control they can attain through the tools used in all the three methods of abuse to perform abusive actions.
The stalking example in Table \ref{abuse_comparison} shows that tech-abuse using smart home devices calls for less effort, lower risk, and more rewards for abusers. 
This makes exploiting smart home devices for abusers in IPV an ideal tool to carry out their harmful activities. 

\subsection{Threat model}
\label{threat_model}

The default trust assumption in homes is problematic and could be abused by intimate partners. 
We assume devices and networks in smart homes are secure from device malfunctions, software bugs, or being compromised with the current security measures. 
Our threat model lies when authenticated users are legitimate, but with ill intentions to abuse devices' features to harm their intimate partners. 
Freed et al. refer to this threat model as UI-bound adversary~\cite{SmartHome_Abuse_Category_1_paper2}.
Attackers (abusers) do not need to be sophisticated, nor delusional. 
They simply use features of the smart home devices legitimately (but maliciously) to carry out abusive actions without the need for programming skills or privilege escalation.
\section{Analytical Framework for SHOT}
\label{phases}

Recent studies have shown that domestic abuse takes different phases.
The phases of abuse depend on the methods that abusers take. 
Walker explains a few phases of traditional abuse where abusers do not use technologies in their actions~\cite{Walker_Traditional_abuse}.
Matthews et al. consider that three phases of abuse are the most appropriate for tech-abuse: physical control, escape, and life apart~\cite{Stories_survivors_IPV}.
In the physical control phase, both abusers and victims of IPV can access shared devices within close proximity. 
In the escape phase, abusers try to retain their access, while victims try to revoke the abusers' access. 
This is the most dangerous phase since abusers normally escalate their aggressiveness before they lose control~\cite{SmartHome_Abuse_Category_2_paper4}. 
The last phase is life apart where victims of IPV try to move on without any contact with their abusers. 
In order for any solution to be effective in combating domestic abuse, whether traditional or tech-based, it needs to address protection in all four phases of abuse, as each phase calls for specific actions. 
Two challenges are missing in recent papers: (1) Who configures smart home devices is a key component in attaining control or preventing control, and (2) Some victims can not recognize the abuse itself or the tools that the abusers use~\cite{AugustLock_Analysis_owner_Capabilities_example,RecognizingAbuse_Traditional,RecognizingAbuse_Traditional_2}.
Therefore, there is a need to redesign the abuse framework to include the missing components in SHOT.

As illustrated in Figure \ref{phases_figure}, we design our framework that consists of four phases of SHOT from the victim's perspective: acquiring smart home devices, recognizing abusive behaviors by intimate partners using these devices, acting upon these behaviors, and living after separation from abusive partners. 
In the acquiring phase, whoever configures the device will have more power to control it compared to other users in the same home.
This is important as the owner account is the decision maker because he or she can manage all users, log usage history, and revoke access~\cite{AugustLock_Analysis_owner_Capabilities_example}. 
In the recognizing phase, victims do not always recognize abusive actions by their partners as abusers normally hide their tools~\cite{RecognizingAbuse_Traditional,RecognizingAbuse_Traditional_2}. 
For instance, recently on the news, a husband confronted his wife over a private conversation that she had with her mom while she was not aware that he was listening~\cite{News_SmartHome_Abuse}.
This example of eavesdropping shows that the IPV victim did not recognize the abuse.
In the acting phase, IPV victims try to reach out for help and record evidence of abuse stealthily to protect themselves.
Lastly, victims and abusers will live separately, and this phase will explore the tools that victims can use to keep their smart home safe. 
Figure \ref{phases_figure} shows the important goals in each phase for each entity involved in IPV.
For example, IPV victims in the recognizing phase aim to look for signs of abuse such as suspicious activities by the smart home devices.
It would be greatly helpful if devices were equipped with features to help victims confirm such suspicions.
Unfortunately, this is not the feasible with current smart home devices.

\subsection{Acquiring a smart home device}
This phase includes purchasing, setting up, and sharing access to smart home devices.
The research community has paid little attention to issues that occur during this phase even though the risk of imbalance in control starts here~\cite{power_Imbalance_Study,European_Cybersecurity_report_imbalanceControlProblem}. 
In a trusting environment, users share passwords to their devices or accounts with their partners~\cite{SmartHome_Abuse_Category_2_paper4}.
This is not an issue unless one partner starts to exploit the established trust.

Parkin et al. have considered a configuration phase in their work \cite{SmartHome_Abuse_Category_2_paper3}. 
However, we believe that the risk exists even before configuring the smart home device.
If users (potential victims) of smart home devices understand accurately what the capabilities of the devices are and what kind of control they offer, they may change their device purchase decision, as changes are easier to make before the purchase~\cite{PurchaseChange}. 
Given the fact that changing behaviors before acquiring the device is possible, we need to consider adding this phase as a crucial component of the abuse phases. 
If the research community puts effort to raise awareness of the devices' susceptibility to exploitation, consumers can make more educated choices. 
This will put pressure on vendors to pay attention to issues in SHOT, and will in turn motivate them to make their devices abuse-resistant.

Recently, a few organizations have attempted to score IoT devices based on their security, privacy, and safety measures~\cite{digital_Standard}. 
The Digital Standards project is run and maintained by the Consumer Report~\cite{consumerReport}. 
The goal is to build standards for IoT devices to help consumers choose the best options for them.
One criterion proposed by the Digital Standards is called ``personal safety.'' It accounts for the measures taken by smart home vendors to mitigate or respond to domestic abuse.
Unfortunately, the corresponding fields on the standard are not filled yet due to a lack of research focus on the matter.
Emami-Naeini et al. have also considered adding a personal safety label. However, more research needs to be conducted to come up with metrics to evaluate devices' capabilities regarding personal safety~\cite{Security_lables}.
We believe that our framework will help define future metrics by showing the different phases of abuse in SHOT and the corresponding device requirements for abuse-resistance. 


The goals in this phase are to limit abusers' excessive control power, educate victims on safety specifications of smart home devices, and assist support services.
First, the research community has proposed a few ways to prevent owners of smart home devices from gaining excessive control power over other users in the same household. 
Device vendors can deploy a setup process that involves other users such as asking the main user to answer whether there are other people living in the same house~\cite{SmartHome_Abuse_Category_2_paper1,power_Imbalance_Study}. 
This also calls for understanding the different levels of technical backgrounds among users. 
One limitation to this approach is that it relies on the honesty of owners to disclose such information, which can be challenging as device owners may pose a threat to the other users in an IPV setting. 
Future research needs to focus on how to enhance the setup process without fully relying on owners' disclosures. 
Second, empowering victims of IPV can be done by providing safety labels on smart home devices~\cite{digital_Standard,Security_lables}. 

Last, support services' goals consist of providing help to victims and holding abusers accountable. These services include front liners, law enforcement, and legal experts. 
Lopez-Neira et al. show that front liners in support organizations suffer from a lack of expertise to deal with SHOT~\cite{SmartHome_Abuse_Category_1_paper1}. 
The authors offer several recommendations such as involving cybersecurity experts to help make safety plans for victims along with front liners at support services.
In Section \ref{ourProposals}, we discuss solutions for current smart home devices. 
We also lay out the desired properties for each phase of the SHOT framework, including the acquiring and recognizing phases.
\subsection{Recognizing abuse in smart homes}

Parkin et al. and Matthews et al. have considered three phases of tech-abuse~\cite{SmartHome_Abuse_Category_2_paper3,Stories_survivors_IPV}.
The first two phases are physical control and escape. 
In this paper, we argue that their approaches miss a very important phase from the victim's perspective. In their approach, physical control means that the abuser still has physical access to the devices in the house. 
The problem here comes from the fact that victims of IPV struggle to recognize abuse~\cite{AugustLock_Analysis_owner_Capabilities_example,RecognizingAbuse_Traditional,RecognizingAbuse_Traditional_2,social_media_IPV}. 
With traditional methods of abuse or tech-abuse in smartphones or computers, organizations have come up with guidelines to help victims recognize patterns or signs of abuse~\cite{NDVH_abuse_signs}. Goulden et al. explain some challenges of living with interpersonal data. 
The authors conduct a few studies to measure accountability and observability leading to the need for new careful designs of smart home technologies to prevent unintended consequences such as tech-abuse~\cite{SmartHome_Abuse_Category_2_paper5}.

The research community has not paid attention to techniques that can assist victims of IPV with recognizing abuse in smart homes. 
Device vendors can develop tools to alarm users of the house with respect to whom these alarms should appear. 
Suppose a smart home device is equipped with safety features to recognize potential domestic abuse. The context-aware device could show warnings --such as playing some sounds to explain some data flows for users in the house-- to warn of remote access~\cite{SmartHome_Abuse_Category_2_paper1,SmartHome_Abuse_Category_1_paper2}.

A context-aware authorization scheme is a great option to help victims of IPV recognize abuse signals. 
Ghosh et al. propose \textit{softAuthZ} that is an authorization scheme in smart homes. 
It incorporates soft security attributes such as the trust level between users and other contextual variables to support authorization management~\cite{softAuthZ_ContextAware}.
IPV victims can use \textit{softAuthZ} to understand the sources of requests and the contextual information.
Moreover, using voice recognition to differentiate between users can help prevent excessive control power that abusers take advantage of~\cite{SmartHome_Abuse_Category_2_paper1}. 
Voice recognition can be used as well to notify the appropriate party if abuse is suspected. 
This calls for a continuous authentication mechanism so that the voice assistant can tell who is sending the commands and who might be abused.
Work like~\cite{voice_assistant_continuousAuthntication} can be utilized to accommodate the needs of distinguishing users.

The goals in this phase are to mitigate abuser's harmful activities, notify victims of signs for potential abusive activities, and assist support services. 
First, implementing context-aware dynamic firewalls can limit abusers' excessive control power in IPV settings~\cite{SmartHome_Abuse_Category_2_paper1}.
Second, empowering IPV victims by making support resources reachable through smart home devices, enhancing data flow visualization  so that users can make sense of activities in their home, and implementing context-aware dynamics firewalls~\cite{SmartHome_Abuse_Category_2_paper1,SmartHome_Abuse_Category_2_paper3}. 
Last, smart home vendors can provide front-liners in support services with guidelines on how to help victims of IPV in SHOT~\cite{SmartHome_Abuse_Category_1_paper1}.

\subsection{Acting upon abuse in smart homes}
This phase is not unique to tech-abuse in smart homes. 
Traditional methods of abuse and tech-abuse have considered this phase of abuse. 
Various legislative bodies and civil rights groups have fought to stop and prevent traditional abuse, and to hold abusers accountable by passing laws, pushing standards, and providing shelter resources. 
For instance, the Colorado state penalizes stalkers by prison sentences up to 10 years~\cite{colorado_law_stalking}. 
These strict laws may deter stalkers in traditional abuse cases.
However, building a legal case with evidence in tech-abuse can be challenging as abusers can find alternative ways to launch their abusive actions~\cite{NY_Times_Tech_Abuse}. 
Recent research has created new tools for victims of IPV to stop their abusers from having access to their phones~\cite{tech-abuse-helpVictims_1,tech-abuse-helpVictims_2}.
Even though the literature has not fully remedied tech-abuse using smartphones and personal computers, adequate resources have been developed for victims~\cite{Tech-abuse-smartphones-computers}.

Compared to the previous approaches of abuse, empowering IPV victims using smart home devices with tools and resources has been limited. 
Some research projects aim to raise awareness and provide proper tools directly to the victims~\cite{SmartHome_Abuse_Category_2_paper1,SmartHome_Abuse_Category_2_paper2,SmartHome_Abuse_Category_2_paper3,SmartHome_Abuse_Category_2_paper4}. 
For example, Parkin et al. analyze two popular smart home devices: \textit{Google Home} and \textit{Amazon Alexa} for their usability in IPV. They find that remodeling the definition of usability in smart home devices is needed to meet the expectations of different entities involved in IPV~\cite{SmartHome_Abuse_Category_2_paper3}.
Others aim to facilitate support services for the victims through local organizations and law enforcement~\cite{SmartHome_Abuse_Category_1_paper1,SmartHome_Abuse_Category_2_paper4,SmartHome_Abuse_Category_3_paper1}.
For instance, authors in~\cite{SmartHome_Abuse_Category_1_paper1} recommend continuously training support services on emerging technologies so that they are well-prepared when fighting SHOT.


It is worth noting that smart home devices empower abusers by default. 
The work in~\cite{SmartHome_Abuse_Category_1_paper1,SmartHome_Abuse_Category_1_paper2,NDVH_abuse_signs} demonstrates that through interviews and workshop involving victims and support services. 
In Section \ref{technical_causes}, we discuss the technical factors that empower abusers over victims in IPV.

One goal of this phase is to assist victims in gathering evidence, hiding their tracks for reaching out to support services, and building a legal case to escape safely. 
Freed et al. show how dangerous this phase can be when victims start to take actions as abusers will escalate their aggressiveness~\cite{SmartHome_Abuse_Category_2_paper4}. 

The other goal is to mediate the help provided by support services to the victim such as facilitating a safe escape plan where abusers should not notice anything.
Islam et al. developed a system that recognizes abuse from the spoken language of victims using Natural Language Processing (NLP) tools~\cite{homeGuard_abuse_recognize_seekHelp}. 
It then evaluates the abuse and reaches out to the nearest support services based on the victim's location. Services might include the police, a nearby hospital, or available lawyers.

\subsection{Living after abuse is identified in smart homes}

\begin{table}[!t]
	\centering
	\caption{Summary of security and safety needs for each phase} 
	\vspace{-8pt}
	\label{tab:techs}
	\scalebox{0.9}{
	\begin{tabular}{|p{8.5cm}|}
		\hline
		
	\rowcolor{Gray}
		\textbf{Acquiring}            \\ \hline
		
		\begin{itemize}
        \item Capabilities of smart home devices should be shown (similar to nutrition labels), including safety measures.
        \item Authentication  mechanisms should consider   multi-users' environments.
        \item Setup processes should consider involving other occupants. 
    \end{itemize}       \\ \hline
    	
    	\rowcolor{Gray}
    	\textbf{Recognizing}            \\ \hline
		
		\begin{itemize}
        \item Implementing continuous authentication to differentiate between users and to include context.
        \item Implementing context-aware access control can help mitigate IPV.
        \item Using nudges to show users data flow and corresponding purposes can raise victms' awareness of abuse risk.
        \item Devices should give equal access power to users in the same home if they are partners.
         \end{itemize}       \\ \hline
    	
    	\rowcolor{Gray}
    	\textbf{Acting}            \\ \hline
		
		\begin{itemize}
        \item Solid security logging can help create evidence and protect victims of IPV.
        \item Vendors should consider adding ways to facilitate reporting domestic abuse if identified.
        \item The research community should aim to develop adversarial models of tech-abuse for different devices. Voice and image recognition or traffic analysis may be sources of data, but user privacy needs to be taken into account.
        \item Access control needs to be reliable and immediate.
    \end{itemize}       \\ \hline
    		
    	\rowcolor{Gray}
    	\textbf{Living}            \\ \hline
		
		\begin{itemize}
        \item Current cybersecurity measures may be good enough. 
        \item There is a need to raise user awareness of cybersecurity hygiene.
        \item Restraining orders need to account for smart home devices.
    \end{itemize}       \\ \hline
	\end{tabular}
	}
	\label{Table_phases_summary_needs}
	\vspace{-10pt}
\end{table}

In this phase, victims seek to establish a new life without their abusers. They may move to a new place, revoke abusers' access to smart homes, or have a restraining order on their abusers. 
Sharing access to smart home devices might get tricky as it is not clear who should ``own" the shared device after separation. 
In addition, living completely independently might be impossible if the victim and the abuser have kids together~\cite{Stories_survivors_IPV}.

It is important to notice that separating procedures take a long time. Victims of IPV need to ensure they are safe during that period~\cite{Seperating_takes_time}.
Transfer of ownership of smart home devices needs to be accounted for in the design stages, so that when victims of IPV want to revoke abusers' access, the process is easy and takes effect immediately~\cite{challenges_safety_smart_Environments}.
In this phase, the abuser should have lost access to the smart home devices, which they used to carry out harmful actions on  towards the victim. The abuser might try to regain access. 
The goal is to teach the victim how to revoke access from the abuser and keep access to the smart home devices private and secure. 
Particularly, victims of IPV need to be knowledgeable about privacy and security best practices to protect themselves~\cite{Stories_survivors_IPV}.

Legally, lawyers of the victims of IPV have difficulties including the access to smart home devices in restraining orders to stop abusers from attempting to regain access~\cite{NY_Times_Tech_Abuse}.
Traditionally, restraining orders account for the victim's home and work buildings.
But, abusers exploiting SHOT do not to be in proximity to carry out harm. 
This adds complexities in writing restraining orders for lawyers and judges as well.
Moreover, security tools need to be effective in terms of timeliness and reliability so that victims can trust the devices to perform as expected~\cite{Access_control_not_working}.
We foresee the need for reliable access control in smart home devices so that revoking requests can be immediate even if usability is slightly compromised. 

Table \ref{Table_phases_summary_needs} summarizes the needs and potential solutions for each phase in our framework.
None of the previous work has considered all the phases in their designs which results in missing abuse-resistant smart home devices.

\section{Technical limitations violate smart home safety}

\label{technical_causes}
In this section, we investigate why current smart home devices empower abusers in IPV from a technical standpoint. We explore how they limit victims' abilities to protect themselves and how they may obstruct support services.
The literature shows that smart home device designers do not consider multi-user environments in the authentication and access control mechanisms, which leads to empowering abusers over victims of IPV~\cite{Access_control_not_working,European_Cybersecurity_report_imbalanceControlProblem}. 

\textbf{The blackbox nature of smart home devices is troublesome.} Litão explains how this blackbox nature inhibits users' abilities to understand the data flow and how this plays a big role in the unequal control power between intimate partners~\cite{SmartHome_Abuse_Category_2_paper1}.
The majority of smart home devices do not offer an interface on the same device. Generally, they use of smartphones for authentication, access control, and privacy settings~\cite{European_Cybersecurity_report_imbalanceControlProblem}. 

\textbf{Recent security research did not consider UI-Adversary in their threat models.}
Slupska conducted a thorough analysis of 44 research papers about security in smart homes.She found that none of them had considered IPV in their threat models~\cite{feminist_Cybersecurity}.
Recent research focused on the same security issues as in personal computers and smartphones. 
Smart homes are different from traditional technology: Social complexities and potential unequal access are extremely important to the safety of users~\cite{European_Cybersecurity_report_imbalanceControlProblem}.

\textbf{Access control in smart home devices is not reliable.}
Janes et al. evaluate 19 popular smart home devices and find that 16 of them struggle to revoke access from previously authorized users even after changing passwords or explicitly revoking their access~\cite{Access_control_not_working}. 
The findings in~\cite{Access_control_not_working} show that these changes are either not enforced at all or do not take effect immediately, making the majority of the smart home devices not reliable when it comes to protecting victims.

\textbf{Smart home devices are not designed for a multi-user environment.}
Recent work discusses why current smart home devices are not designed for multiple users. The authors of ~\cite{multi-user-problem-smart-home,Rethinking_Access_control_Authnetication} conducted user studies to qualitatively analyze users' needs in smart homes. 
The findings call for flexible authentication and access control. 
This flexibility helps consider the social relationships of users in smart homes. 
Some even call for not implementing any authentication process for people who are within proximity, as is the case for traditional home appliances. 
Others call for a dynamic design where users can have full control with facilitation to add and remove access smoothly. 
Incorporating flexibility in smart home devices by giving users more options to control is not trivial, as many research projects have not been able to balance convenience and privacy~\cite{multi-user-problem-smart-home,Too_many_Notifications}.

\textbf{Smart homes call for more technical background.}
Smart homes are similar to enterprise networks more than personal computers or smartphones.
The expertise and workforce that enterprises have to manage their networks are not available for smart home users~\cite{European_Cybersecurity_report_imbalanceControlProblem}.
The case in smart homes is that some users might have more knowledge than others and this uneven power is caused by inappropriate design in the first place.

\begin{table*}[!t]
	\centering
	\caption{Summary of proposed solutions in the literature to combat SHOT} 
	\vspace{-9pt}
	\label{tab:techs}
	\scalebox{0.9}{
	\begin{tabular}{|p{3cm}|p{0.5cm}|p{5cm}|p{4.4cm}|p{0.9cm}|p{1cm}|}
	    \hline
	    \rowcolor{Gray}
	    
		\textbf{Solution}    &  \textbf{In} &  \textbf{Downsides}&  \textbf{Objective}&  \textbf{Can victims act?}&  \textbf{Short or long term?}       \\ \hline
		
	 Involving cybersecurity experts in designing smart home devices&\cite{SmartHome_Abuse_Category_1_paper1,SmartHome_Abuse_Category_2_paper4,SmartHome_Abuse_Category_2_paper6}&   Vendors need more time and budget before launching their products.&  
	     Ensuring applicability and reliability of security features \textit{and} providing plans for front liners
	
&  No&  Long term       \\ \hline

 Adding labels to smart home devices similar to the nutrition labels&\cite{Security_lables}& 
      Abusers might not buy devices with safety features \textit{and} vendors need more time and budget before launching their products.
  &  
	     Raising awareness for victims \textit{and} pushing for standardizing safety in smart home devices

&  Yes&  Long term       \\ \hline

 Deleting usage history periodically&\cite{SmartHome_Abuse_Category_2_paper4,Stories_survivors_IPV}& 
      
          Might help abusers hide their abuse and gaslight victims~[12] \textit{and} victims might lose evidence when building a case~\cite{SmartHome_Abuse_Category_2_paper4}.
   
  &  
	     Protecting victims by hiding their escape plans
	    
&  Yes&  Short term       \\ \hline

Making usage history permanent  &\cite{IBM_tech_abuse_design}& 
      
          Victims struggle to hide their escape plans for victims \textit{and} personal privacy for normal users might be compromised.
       
  &  
	     Helping hold abusers accountable \textit{and} victims can have evidence when building a case~\cite{Stories_survivors_IPV,SmartHome_Abuse_Category_2_paper4}
	 
  &  No &  Short term       \\ \hline

Providing emergency access to support services to intervene in case of abuse &\cite{SmartHome_Abuse_Category_2_paper6}& 
      
          Making decisions on behalf of users is not good practice~\cite{IBM_tech_abuse_design} \textit{and} personal privacy for normal users might be compromised.
  &  
	     Helping victims reach out to support services
	      &  No &  Long term       \\ \hline

Making devices distinguish users through voice recognition &
\cite{SmartHome_Abuse_Category_2_paper1,SmartHome_Abuse_Category_2_paper6}& 
      
          This leads to less controllable devices~\cite{Access_control_not_working} \textit{and} works only with voice assistants.

  &  
	     Protecting users' data from other users
	    
&  No &  Long term       \\ \hline

Providing Clinical Computer Security for victims of IPV &\cite{tech-abuse-helpVictims_2}.& 
      Solution does not consider acquiring and living phases \textit{and} is hard to scale
  &  
	     Providing personalized help to victims \textit{and} keeping personnel up-to-date
	 
 &  Yes &  Long term       \\ \hline

	\end{tabular}
	} 
	\label{literature_solutions}
	\vspace{-15pt}
\end{table*}


\textbf{Collecting evidence to make legal cases is hard in smart homes.}
If victims identify abuse and want to make a legal case, legal resources are limited~\cite{NY_Times_Tech_Abuse}.
Abusers exploit the excessive power they attain in most smart home devices by deleting usage histories when they gaslight their partners~\cite{SmartHome_Abuse_Category_1_paper2,IBM_tech_abuse_design}.
If abusers can not change the usage history without their partner's knowledge, this can help prevent gaslighting.
However, this same feature can be harmful when victims try to hide their awareness of abuse and their communication with support services~\cite{SmartHome_Abuse_Category_2_paper4,Stories_survivors_IPV}.

\textbf{Lack of adequate training for support services.}
Support services need up-to-date training to keep up with emerging tools that abusers use~\cite{SmartHome_Abuse_Category_2_paper4}.
Front liners are the first point of contact for victims in IPV cases. 
They always need to be equipped with the proper tools to help victims and connecting them with available resources.
The lack of expertise and training in support service providers leaves victims helpless.

\textbf{Lack of standards and regulations that account for IPV in smart homes.}
There are no known regulations that consider SHOT that vendors can follow.
The reason behind the absence of regulations may be related to the lack of standards.
Standards by themselves can not enforce vendors to change their designs. However, standards are usually what regulations refer to when writing legislation and guidance documents~\cite{BSI_Standards_regulations}.
So if standards are created, regulations will likely be developed and enforced.
Writing standards is not trivial and requires a lot of continuous collaborations between stakeholders in the smart home industry.
Piasecki et al. review cybersecurity standards and find that they do not consider the \textit{UI-adversary} in their guidelines and recommendations~\cite{review_standards_IoT}.
In addition, IoT Security Foundation, a leading influencer in the IoT industry, surveyed the IoT industry for basic security adoptions and found that the majority of the IoT vendors were still not compliant with basic security hygiene~\cite{IoT_security_foundation}. 
This shows that the industry is still behind on basics.
Making smart home devices abuse-resistant is a step further. We do not foresee this being adopted in the near future with the current situation despite all the high risks associated with SHOT.

Smart home technology vendors do not intentionally make their devices as tools for domestic abuse. There is in fact great evidence that smart home devices are so beneficial for several reasons such as increasing safety and saving energy.
For instance, the Los Angeles Police Department has reported that the use of Ring doorbells had reduced crimes in some of Los Angeles neighborhoods by 55 percent in seven months~\cite{Ring_Example_CrimeLess}. 
However, smart home technology vendors have not paid attention to the IPV threat. 
Lack of diversity of the design teams, poor usability of the devices, inadequate research of the IPV threat, and system malfunction or unexpected functions are the main reasons for smart home devices being an abuse vector.

\section{State of the art recommendations in SHOT}
\label{summary_solutions}

Recent research and government efforts related to domestic abuse fall in three categories: (1) restraining abusers, (2) developing resources for victims to raise awareness about abuse and facilitate communications with support services, and (3) training personnel involved in countering IPV.
These measures generally help hold abusers accountable and empower victims to protect themselves. 
However, we see a lack of similar initiatives in combating IPV in smart homes~\cite{SmartHome_Abuse_Category_1_paper1}.
In this section, we explore the efforts of the research community to combat IPV that exploit smart home technologies.

Our goal is to compare the proposed solutions and identify similarities, differences, and future directions.
Table \ref{literature_solutions} manifests our summarized list of recent solutions and our assessment of their potential and limitations. 
It also illustrates similarities and differences as well as conflicts in different phases.
It is worth noting that the recommendations in the literature are mostly directed to vendors or support services generically.
Freed et al. discuss the issue of a lack of actionable instructions~\cite{SmartHome_Abuse_Category_2_paper4}.
We do not provide an exhaustive list of the proposed solutions due to the page limit for the paper.
We selected a sample of solutions highlighting the lack of complete, transparent, private, and reliable solutions in the market.

We see conflicting solutions in Table \ref{summary_solutions} because these solutions were designed without consideration of the phases of abuse in smart homes. 
An effective solution to one phase might not work for another phase.
Our framework will assist future designs to prevent conflicting solutions and design a solution that works in all the phases.
Another reasonable assumption is that devices will have modes of operations. So that if abuse is suspected, the device mode will switch to an abusive mode where priorities and functionalities change.
For example, the SmartThings hub has modes of operations such as home, away, and night~\cite{smartthings}.
Modes of operation can include a safety mode, which can be triggered by abusive behaviors.
Future research can address the following questions: (1) What could trigger an abusive mode?, (2) Are these triggers preventable?,
(3) Who should be contacted for help?, and
(4) Does the abusive mode affect user experience in non-abusive environments?

\section{Desirable Proprieties for Abuse-resistant smart homes}
\label{ourProposals}
After revising recent proposals and recommendations in Section \ref{summary_solutions}, it is clear that desirable properties are missing for developing abuse-resistant devices. 
Based on our framework and analysis of the literature in Section \ref{phases}, we recommend a list of \textbf{desirable properties for designing abuse-resistant smart home devices below.}
\begin{itemize}
\item \textbf{Completeness:}  A complete solution accounts for all phases of abuse. Technical, legal, and ethical perspectives need to be considered for a complete solution.
\item \textbf{Transparency:} This helps create context and ideally will help raise awareness and prevent abuse~\cite{Access_control_not_working,IBM_tech_abuse_design}.
\item \textbf{Equal access for partners:} Devices should be designed for multiple users by default. 
For example, when equal partners share a smart home, both should have equal access power to the devices. 
This also calls for understanding the technical gaps between users, so that vendors can design appropriate devices even for less tech-savvy users.
\item \textbf{Privacy-preserving safety features:} Modes of operations should account for abusive triggers and environments. 
This helps maintain utility for non-abusive environments.
\item \textbf{Reliability:} Smart home devices should prioritize safety over convenience.
For example, revoking access should be considered as an important low-latency request. 
This also can help victims create a legal case when abuse is identified.
\end{itemize}

\section{Conclusion}
\label{conclusion}
Designing smart home devices with an assumption of absolute safety in homes is problematic as victims of IPV suffer in their homes, and escape plans are the only safety boat for them.
In this paper, we articulated the problem of IPV in smart homes and how it is unique.
We then proposed a framework that covers all phases of abuse in smart homes and their corresponding needs and challenges.
We explored the technical reasons that empowered abusers more than victims in current smart home devices. We discussed the recommendations offered by the research community with emphasis on their potentials and limitations.
Finally, we provided a list of desirable properties of ``abuse-resistant" smart home devices.

\end{document}